\documentclass[sn-mathphys-num]{sn-jnl}

\usepackage{lipsum}
\usepackage{graphicx}%
\usepackage{tabularx}
\usepackage{multirow}%
\usepackage{amsmath,amssymb,amsfonts}%
\usepackage{amsthm}%
\usepackage[mathscr]{eucal}
\usepackage[table]{xcolor}
\usepackage{bm}
\usepackage{textcomp}%
\usepackage{manyfoot}%
\usepackage{booktabs}%
\usepackage{algorithm}%
\usepackage{algorithmicx}%
\usepackage{algpseudocode}%
\usepackage{listings}%
\usepackage{natbib}
\usepackage{array}
\usepackage[normalem]{ulem}
\newcolumntype{P}[1]{>{\centering\arraybackslash}p{#1}}

\theoremstyle{thmstyleone}%
\theoremstyle{thmstyletwo}%
\theoremstyle{thmstylethree}%
\begin{document}

\title[Article Title]{Classification of SARS-CoV-2 Variants through The Epistatic Circos Plots with Convolutional Neural Networks}

\author[1]{\fnm{Bo} \sur{Jing}}
\author[1]{\fnm{Kai-Rui} \sur{Zhang}}
\author*[1,2]{\fnm{Hong-Li} \sur{Zeng}}\email{hlzeng@njupt.edu.cn}

\author*[3]{\fnm{Erik } \sur{Aurell}}\email{eaurell@kth.se}

\affil[1]{\orgdiv{School of Science}, \orgname{Nanjing University of Posts and Telecommunications}, \orgaddress{ \city{Nanjing}, \postcode{210023}, \country{China}}}

\affil[2]{\orgdiv{National Laboratory of Solid State Microstructures}, \orgname{Nanjing University}, \orgaddress{\city{Nanjing}, \postcode{210093}, \country{China}}}

\affil[3]{\orgdiv{Department of Computational Science and Technology}, \orgname{AlbaNova University Center}, \orgaddress{\city{Stockholm}, \postcode{SE-106 91},  \country{SWEDEN}}}

\abstract{
The COVID-19 pandemic has profoundly affected global health, driven by the remarkable transmissibility and mutational adaptability of the SARS-CoV-2 virus. Although five variants of concern, Alpha, Beta, Gamma, Delta, and Omicron, have been identified, the classification task in this study is formulated using four classes: Alpha, Delta, Omicron, and Else, reflecting the sequence availability and temporal coverage of the dataset. Here, we develop an integrative framework that combines direct coupling analysis (DCA), Circos-based visualization, and convolutional neural networks (CNNs) to characterize lineage-specific epistatic signatures from large-scale SARS-CoV-2 genomic sequences. DCA-inferred pairwise mutational couplings were transformed into Circos images, which were then used as inputs for CNN-based classification models.
The proposed framework achieved robust variant classification, with the best-performing model reaching a weighted-average F1-score of $98.68\pm 0.75\%$ and an AUC close to 1.}

\keywords{SARS-CoV-2, Convolution neural network, Direct coupling analysis (DCA), Epistasis, Circos plots}


\maketitle

\section{Introduction}

The COVID-19 pandemic, caused by the severe acute respiratory syndrome coronavirus 2 (SARS-CoV-2), has posed an unprecedented global health crisis since its emergence in late 2019. The virus exhibits remarkable transmissibility and mutational adaptability, which have enabled it to spread rapidly and persist across diverse populations. Over the pandemic period, 
several major variants of concern (VOCs) including Alpha, Beta, Gamma, Delta, and Omicron have been identified by WHO, each characterized by distinct combinations of mutations associated with enhanced infectivity, immune escape, or altered pathogenicity~\cite{covid19, harvey2021sars}. Understanding the evolutionary mechanisms underlying these adaptive features is critical for developing effective surveillance strategies, guiding vaccine updates, and anticipating future viral dynamics.

The advent of high-throughput sequencing technologies has generated a large amount of viral genomic data. Until August 2022, the GISAID repository contained more than 12.95 million high-quality SARS-CoV-2 genome sequences~\cite{gisaid}. This vast dataset offers an exceptional opportunity to study the evolutionary trajectories of SARS-CoV-2 at the molecular level. Beyond identifying single mutations or lineage-defining substitutions, researchers have increasingly focused on the epistatic interactions—non-additive effects arising from interdependent mutations—that shape the virus’s adaptive landscape. Such epistasis can modulate viral fitness, alter host–pathogen interactions, and influence antigenic drift, thereby serving as a key driver of viral evolution~\cite{moulana2022compensatory, witte2023epistasis}.

To uncover these epistatic relationships, various computational frameworks have been developed, among which Direct Coupling Analysis (DCA) has gained particular prominence. Originating from statistical physics, DCA infers residue–residue coevolution by estimating direct information couplings between positions in multiple sequence alignments~\cite{Ekeberg-2013a, Ekeberg-2014a, Weigt2009high}. Recent applications of DCA to SARS-CoV-2 have revealed complex epistasis of co-mutations through the genomic sequences~\cite{Zeng2020PNAS,CRESSWELLCLAY20211,ZENG2022PRE,Rodriguez-Rivas2022,Zeng2025MPL_DCA}. The results are often visualized as Circos plots~\cite{krzywinski2009circos}, which display pairwise genetic couplings across the viral genome. These circular representations allow researchers to intuitively inspect patterns of co-variation and potential functional dependencies. However, as the scale of genomic data and the number of variants continue to grow, manual inspection and traditional statistical evaluations of these Circos plots become increasingly inefficient.

To address this challenge, there is a growing need for automated, scalable, and data-driven approaches capable of detecting and classifying complex epistatic patterns embedded in visual data. Notably, Circos plots that represent genome-wide epistatic interactions can be regarded as structured images in which spatial arrangements and link densities encode biologically meaningful relationships. Therefore, identifying variant-specific epistatic signatures becomes essentially a pattern recognition problem in a high-dimensional visual space.

Recent advances in deep learning, particularly convolutional neural networks (CNNs), have demonstrated extraordinary success in image recognition, natural pattern extraction, and multidimensional data analysis. CNNs are designed to hierarchically learn spatial features from raw visual inputs, allowing them to capture both local and global dependencies without the need for handcrafted feature engineering~\cite{lecun2015deep, krizhevsky2012imagenet}.  
In bioinformatics, deep convolutional and representation-learning architectures have been successfully applied to problems such as protein structure prediction, chromatin interaction mapping, and molecular phenotype classification~\cite{senior2020improved, schreiber2020avocado, zhou2015predicting}.

Inspired by the above insights, we propose to leverage CNNs to analyze Circos plots derived from DCA method over the genomic sequences of SARS-CoV-2~\cite{ZENG2022PRE}, thereby enabling the automated identification of genomic variants through their epistatic signatures. By treating Circos plots as structured visual representations of co-mutation networks, CNNs can learn the features of the spatial configurations corresponding to different evolutionary lineages. Here we constructed a dataset of 1,984 Circos plots encompassing five major VOCs and developed a tailored CNN classification framework to classify the variants. Here the classification task focuses on Alpha, Delta, Omicron, and an aggregated ``Else" category. This is motivated by the limited sample size and short temporal prevalence of Beta and Gamma variants in the available dataset, which makes it difficult to train statistically robust classifiers for these classes individually. The trained model achieved a classification accuracy of 98.68\%, demonstrating its ability to extract and generalize the complex spatial and statistical patterns encoded in the Circos visualizations.

Our results indicate that CNNs can serve as powerful tools for recognizing structural signatures associated with  higher-order genetic dependencies, offering a novel avenue for rapid screening and profiling of complex evolutionary patterns from large-scale genomic data. Furthermore, this approach bridges the gap between statistical physics-based evolutionary modeling and artificial intelligence, providing a scalable and interpretable framework for exploring viral coevolution. By integrating DCA-based inference with deep learning-based visual recognition, we aim to advance our understanding of the complex evolutionary landscapes of SARS-CoV-2 and potentially other rapidly evolving RNA viruses, providing a hypothesis-generating platform to guide future experimental studies.

The remainder of this paper is organized as follows. Section~\ref{sec:related_work} presents the related work using wider types of AI models on sequence analysis,
DCA models are introduced above, or, as appropriate, in later section.
Section~\ref{sec:materials_methods}
describes the construction of the dataset, DCA approach and  CNN architectures adopted for image classifications. Section~\ref{sec:results} provides the experimental results and evaluation criteria. Section~\ref{sec:discussion} and~\ref{sec:conclusion} discuss and conclude the results in our work and outlines perspectives for future researches.

\section{Related work} \label{sec:related_work}
Technical background details on the DCA method (which we use in this work) are presented separately and in more detail in Materials and Methods, while key references in this line of research are cited above in Introduction. We have chosen to also place a discussion of the connection of DCA to Quasi-Linkage Equilibrium in the context of Materials and Method.
Here we survey the relevant wider literature
on applying AI methods to inference of biological properties. As goes without saying, this literature is very large, and the following should mostly be taken as indications of the convenient entry points.

First the pioneering contribution of the Circos software, soon to be 20 years ago, has to be acknowledged again~\cite{krzywinski2009circos}.
We use this tool extensively in this paper.
Second, before the AlfaFold revolution, many attempts were made to leverage deep learning for biological classification and analysis, some examples being
\cite{Angermueller2016,Ren2020,Tampuu2018}.
Also computationally simpler method were applied extensively, see e.g. \cite{Ren2017}. From the perspective of DCA and DCA-like methods, attempts were made 
in \cite{Skwark2017} to identify combinations of mutations leading to antibiotic resistance in 
\textit{Streptococcus pneumoniae} and 
\textit{Streptococcus pyogenes}.
The field was transformed by AlphaFold
in its many successive variants and extensions, showing that 
protein 3D structure, the most important determinant of protein fitness, can be accurately inferred from homologous protein sequences (protein families)~\cite{Jumper2021,Mirdita2022,Cheng2023,Lin2023,Madani2023,Ferruz2022}.
Many later contributions aim to extend deep learning
methods to structure sets of genome sequences, to separate human from viral sequences and many analogous tasks 
\textit{cf.}~\cite{Ji2021,Zhou2024,Dalla-Torre2025,Rancati2025.12.12}.
Such investigations also extend to SARS-CoV-2 data
to elucidate its evolutionary dynamics,
to forecast dominance of lineages,
and to specifically understand the Spike protein, its variability and effects of these
\cite{Zvyagin2022,Dhodapkar2023,Wang2023,Rancati2024,Rancati2025,Symons2025,Lamb2026}.

A more comprehensive discussion on fitness effects of mutations to SARS-CoV-2 proteins can be found in
\cite{Bloom2023}, while aspects of generative landscapes 
impacting on functional diversity in protein sequence space,
viral immune escape and 
language models to explore viral fitness landscapes
can be found in \cite{Zeigler2023,Ito2025,Huot2}.

\section{Materials and Methods} \label{sec:materials_methods}

\subsection{Data Collection}\label{sec3.1}

All genome sequences analyzed here were retrieved from GISAID \cite{gisaid, Lopez-Rincon2021}, which has evolved into one of the most important repositories for SARS-CoV-2 genomics. It has continuously aggregated vast amounts of high-quality genome sequences submitted by research institutions and public health laboratories worldwide, thereby providing an indispensable foundation for large-scale studies on viral evolution, mutation tracking, and epidemiological surveillance. To date, millions of complete SARS-CoV-2 genome sequences have been deposited in GISAID, providing an unparalleled resource for tracking viral evolution. However, due to a temporal delay between sample collection and sequence submission \cite{kalia2021lag}, the database reflects viral dynamics with a slight time lag. To ensure robust temporal representation, we therefore compiled as comprehensive a dataset as possible, covering diverse sampling periods and geographic regions.

To keep the analysis consistent with \cite{ZENG2022PRE}, we employ the same dataset covering the period from late 2019 to August 23, 2022, as the number of new genome submissions to GISAID declined significantly thereafter. We collected 5,465,988 high-quality, full length genome sequences (approximately 30 kbp each)  from GISAID for analysis. To ensure temporal consistency, the sequences were stratified on a daily basis, yielding data spanning 962 consecutive days. The number of entries per day reflects the volume of genome submissions to GISAID as of the above date. 

Each entry in the GISAID database is accompanied by detailed metadata, including gender, age, geographic region, and viral lineage, which provide valuable contextual information for the following classification and comparative analyses.

\begin{figure*}[!ht]
\centering
\includegraphics[width=0.65\textwidth]{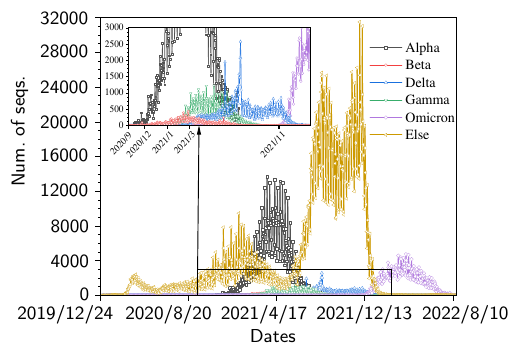}
\caption{Temporal distribution of SARS-CoV-2 genome sequences by variants.
The main panel shows the daily number of sequences collected from the GISAID database between December 2019 and August 2022, classified into the major variants of concern (Alpha, Beta, Gamma, Delta, and Omicron) and other minor lineages (Else). Distinct temporal waves correspond to the global emergence and dominance of each variant: Alpha surged in early 2021, Delta dominated mid-2021, and Omicron replaced all previous variants after late 2021. The inset highlights the early coexistence of Alpha, Beta, and Gamma variants. The sequencing volume declines after mid-2022, reflecting reduced global sequencing activity and fewer new submissions to GISAID. } 
\label{Fig.1} 
\end{figure*}

\subsection{Data Preprocessing}\label{sec:data-preprocessing}

This section details the data preprocessing framework used in this study. Specifically, raw SARS-CoV-2 genomic sequences retrieved from GISAID were subjected to systematic cleaning, alignment, and quality filtering to construct high-confidence multiple sequence alignments (MSAs). Subsequently, the Direct Coupling Analysis (DCA) method was applied to infer epistatic interactions between genomic loci, which were then visualized using the Circos software~\cite{krzywinski2009circos}. The structured visual epistasis is required for downstream deep-learning-based classification and variant identification.

Fig. \ref{Fig.1} illustrates the temporal distribution of SARS-CoV-2 genome sequences across different variants between December 2019 and August 2022. The number of sequences exhibits clear temporal clustering corresponding to the global emergence of distinct variants. The Alpha wave peaked in early 2021 and was subsequently replaced by Delta in mid-2021, while Omicron became dominant after November 2021, producing the highest sequencing counts. In contrast, Beta and Gamma remained relatively limited throughout the period. The Else category represents other minor or unclassified lineages that persisted at low levels. The gradual decrease in sequence submissions after mid-2022 indicates a decline in global genomic surveillance and GISAID data reporting. The temporal distribution of variations highlights the differing evolutionary dynamics and data availability among the major variants.

The genomic data were classified according to variant type and sampling date using a MATLAB-based preprocessing pipeline, which generated sub-datasets for each strain and time interval. Since DCA requires a sufficient sample size to reliably infer epistatic interactions, all sub-datasets containing 150 or fewer sequences were excluded. With such filtering, a total of 906,292 Alpha sequences (259 days), 119,855 Delta sequences (259 days), 414,728 Omicron sequences (224 days), and 4,016,078 Else sequences (700 days) were retained for downstream analysis.

\subsubsection{Sequence Alignment and loci filtering}

All remained raw genomic sequences within each variant group were aligned using multiple sequence alignment (MSA) to ensure positional correspondence across nucleotide sites. Here the online server MAFFT \cite{katoh2019mafft,kuraku2013aleaves} was used to perform the alignment with ``Wuhan-Hu-1" \cite{wu2020new} serving as the reference sequence. Such manipulation not only speeds up the alignment but also reduces the burden on computational resources.  

With the obtained MSAs, each sub dataset can be represented as a large matrix $S=\{ \sigma^n_i|i=1,2,...,N\}$, where $N$ denotes the number of aligned genomic sequences in the sub dataset and $L$ the number of loci. Here $L=29,903$ which has the same length with the reference sequence.  Each loci $\sigma^n_i$ of the matrix $S$ can take six possible states: 
four main nucleotides (A, G, C, T), an ``unknown nucleotide" (N), or an alignment gap (-) introduced to handle nucleotide deletions or insertions~\cite{Zeng2020PNAS}. For computational convenience, these states (“–NAGCT”) were numerically encoded as (0–5) to facilitate inferring processing. 

To remove non-informative positions and reduce redundancy in the epistatic inference, we applied a two-stage filtering strategy controlled by a threshold parameter $\phi$ (expressed as a percentage). First, for each column in the MSA, if the frequency of the most dominant nucleotide exceeded $\phi$, the corresponding locus was considered overly conserved and removed. Second, for each sequence, if the frequency of any single nucleotide exceeded $\phi$, or if the total fraction of canonical bases (A, C, G, and T) was less than or equal to
$1-\phi$, that sequence was excluded.

\begin{figure*}[!ht]
\centering 
\includegraphics[width=0.8\textwidth]{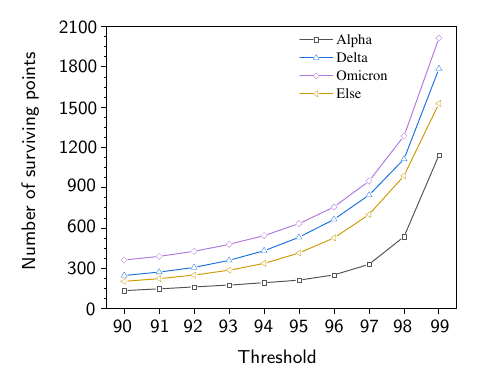}
\caption{Effect of filtering threshold on the number of surviving loci for four variant categories. With the threshold $\phi$ increases from 0.90 to 0.99, the number of surviving loci rises steadily, reflecting the progressive removal of highly conserved sites. Among all variants, Omicron consistently retains the largest number of loci, followed by Delta, Else, and Alpha, indicating higher overall sequence variability in Omicron and Delta compared to Alpha. A near-exponential growth in surviving loci is observed when $\phi > 0.96$, demonstrating the sensitivity of the dataset size to the chosen conservation threshold.} 
\label{fig:retained_loci} 
\end{figure*}

This filtering process effectively eliminated loci with low mutational variability and sequences with poor information content, thereby improving the robustness and computational efficiency of subsequent DCA-based epistasis analysis. The variation in the number of retained loci under different thresholds for each viral variant is illustrated in Fig.~\ref{fig:retained_loci}.
The choice of the filtering threshold $\phi$ plays a critical role in balancing data completeness and sequence variability. A lower threshold removes highly conserved loci, thereby focusing the analysis on positions with greater mutational potential, whereas an excessively low threshold may discard too many loci and reduce the effective sequence coverage. Conversely, a higher threshold retains more data but increases redundancy and noise from invariant sites, which can obscure the detection of genuine epistatic couplings.

To evaluate the impact of $\phi$, we systematically tested several threshold values ranging from 0.90 to 0.99 and analyzed the number of surviving loci for each SARS-CoV-2 variant (Fig.~\ref{fig:retained_loci}). The number of retained loci increases monotonically with $\phi$, indicating that a higher threshold selectively removes conserved genomic positions and retains those exhibiting greater sequence diversity. Among the four analyzed variants, Omicron exhibits the largest number of surviving loci across all thresholds, suggesting a higher degree of nucleotide variability, while Alpha remains the most conserved. The curves also show a pronounced rise beyond $\phi = 0.96$, reflecting the nonlinear effect of the threshold on dataset size and emphasizing the importance of selecting an optimal $\phi$ for balancing data quality and sample quantity. To ensure sufficient sample size while maintaining data quality, we combined the datasets obtained under the thresholds $\phi = 0.95$ and $0.96$, providing an optimal compromise between data completeness and sequence diversity. This combination does not introduce redundancy at the sample level, as the same sequences are represented in different feature spaces, capturing complementary patterns of genomic variation and epistatic interactions. This multi-threshold filtering strategy can be interpreted as a form of multi-resolution representation of genomic variability, enabling the model to capture both strongly and moderately variable interaction patterns.

\subsubsection{Direct Coupling Analysis}
The Direct Coupling Analysis (DCA), otherwise alternatively referred to as maxentropy methods, inverse statistical mechanics or model inference in an exponential family, means to learn a probabilistic model of the Gibbs-Boltzmann type from data, and then to use the parameters of this model to predict a quantity of interest.
When the model is limited to linear and quadratic terms the probability distribution is equivalent to the equilibrium distribution of an Ising or Potts model, and can be written 
\begin{equation}\label{eq:ML}
P(\bm{\sigma})=\dfrac{1}{Z}\exp\left\{\sum_i h_i(\sigma_i)+\sum_{ij} J_{ij}(\sigma_i,\sigma_j)\right\},
\end{equation}
In above $\bm{\sigma}$ denotes a $L$-dimensional vector of data,
the parameters $h_i(\sigma_i)$ encode
effects which only depend on the data ($\sigma_i$) at one position $i$, and 
$J_{ij}(\sigma_i,\sigma_j)$ encode effects which depend on the data ($\sigma_i,\sigma_j$) at two positions $i$ and $j$.
The total data then consists of $N$ data vectors, typically assumed to be independent draws of the same probability distribution, such that the total probability is the product of $N$ functions as in 
\eqref{eq:ML}, where the argument of the $r$'th term in the product is the $r$'th draw $\bf{\sigma}^{(r)}$:

\begin{equation}\label{eq:ML-r}
P\left(\{\bm{\sigma}^{(r)}\}_{r=1}^N\right)=
\prod_{r=1}^N
P\left(\bm{\sigma}^{(r)}\right)
=
\dfrac{1}{Z^N}\exp\left\{
N\sum_{i,s} \left<h_i(s)\right>+N\sum_{ij,ss'} 
\left<J_{ij}(s,s')\right>\right\}
\end{equation}
where
$\left<h_i(s)\right>$ stands for $\frac{1}{N}\sum_{r=1}^N
h_i(s)\,
\mathbf{1}_{\sigma_i^{(r)},s}$ and
$\left<J_i(s,s')\right>$ stands for $\frac{1}{N}\sum_{r=1}^N
J_{ij}(s,s')\,
\mathbf{1}_{\sigma_i^{(r)},s}\mathbf{1}_{\sigma_j^{(r)},s'}$.
The original and still most important instance of DCA is learning a model of a protein family, where each
$\bf{\sigma}^{(r)}$ stands for the amino acid sequence of protein $r$, and the arity of each argument is $20$ (number of amino acids), or more often $21$ (including a gap variable in a multiple squance alignment)~\cite{Weigt2009,Morcos2011}, for extensions to protein complexes see 
\cite{Weigt2009high,Hopf2014}.
Large values of $J_{ij}$ (evaluated in a suitable norm) are then good predictors of spatial proximity in the protein structure or between proteins.
DCA has been reviewed multiple times from the algorithmic point of view, see e.g. \cite{Nguyen2017,Cocco2017}.   

The underlying principle of why DCA works, \textit{e.g.} why the terms in the probabilistic model is a better predictor than straightforward correlations has been much debated \textit{cf.}
\cite{Aurell2016,vanNimwegen2016}. Given the diverse nature of biological sequences to which DCA has been applied, pertaining to phenomena developing on very different time scales, there may be different underlying mechanisms in different cases. In particular, DCA may not always be a reasonable inference procedure if actually the underlying distribution is not close to \eqref{eq:ML}.  

In population genetics, the quasi-linkage equilibrium (QLE) framework provides a statistical description of weakly interacting loci under conditions of moderate recombination and weak selection. In this regime, linkage disequilibrium (LD) between loci can be approximated by a linear response to the underlying epistatic interaction coefficients, allowing the genotype distribution to be expressed as a perturbation around an uncorrelated (product) distribution
~\cite{Kimura1965,NeherShraiman2009,NeherShraiman2011,gao2019dca,Zeng2020PRE}. Mathematically, the distribution of individuals over genotypes in a population are then of the Gibbs-Boltzmann form, \textit{i.e.} as in \eqref{eq:ML}.
However, in contrast to other approaches where the parameters $h_i$ and $J_{ij}$
are directly interpreted as fitness effects, in the setting of QLE they are derived compound parameters expressible in terms of fitness effects and mutation and recombination rates \cite{Dichio_2023}. For instance, the relation between $J_{ij}$ and epistatic fitness parameters $f_{ij}$ depends parametrically on mutation rates, and is different at high and low mutation rates~\cite{Zeng2020PRE}.
Furthermore, at sufficiently low recombination rates the distribution of individuals over genotypes is not of the exponential type but closer to a mixture model, a phenomenon referred to as clone competition \cite{NeherShraiman2009,NeherShraiman2011,Neher2013}.  
Although QLE hence simplifies the biological reality by assuming weak selection and moderate recombination, it provides a practical statistical foundation for extracting meaningful co-evolutionary information from large-scale genomic datasets.

Here we employed the pseudo-likelihood maximization (PLM) approach, a computationally efficient variant of DCA, to infer probabilistic relations between loci from aligned MSA \cite{besag1975statistical, ravikumar2010high, aurell2012inverse,Ekeberg-2013a,Ekeberg-2014a,gao2019dca}. The PLM method estimates the coupling parameters by maximizing the conditional probability of each site given all other sites in the alignment. For a Potts model with multiple states ($q > 2$), the conditional probability of site $i$ takes the form:
\begin{equation}\label{eq:PLM_conditinal_dist}
P(\sigma_i|\bm\sigma_i) =\dfrac{\exp\left(h_i(\sigma_i)+\sum_{j\neq i} J_{ij}(\sigma_i,\sigma_j)\right)}{\sum_{\bm{u}}  \exp\left(h_i(u)+\sum_{j\neq i} J_{ij}(u,\sigma_j)\right)},
\end{equation}
with $u=\{0,1,2,3,4,5\}$ as the possible states of $\sigma_{i}$. 
For data consisting of $N$ sequences PLM amounts to separately maximizing $L$ functions   
\begin{equation}\label{eq:pesudo_condi_log_likelihood}
\begin{aligned}
 PL_{i}(h_{i},\{J_{ij}\})
&=\dfrac{1}{N}\sum_{\substack{s}} h_{i}(\sigma_{i}^{(s)})\\
&+\dfrac{1}{N}\sum_{\substack{s}}\sum_{\substack{j\neq i}} J_{ij}(\sigma_{i}^{(s)},\sigma_{j}^{(s)})\\
&-\dfrac{1}{N}\sum_{\substack{s}}\log\sum_{\substack{u}}\exp\left(h_{i}(u)+\sum_{\substack{j\neq i}}J_{ij}(u,\sigma_{j}^{(s)})\right)
\end{aligned}
\end{equation}
with $s$ labeling the sequences from 1 to $N$. An adjustment procedure is applied afterwards to reconcile the two in principle different values obtained for $J_{ij}$ from maximizing respectively $P(\sigma_i|\bm\sigma_i)$ or $P(\sigma_j|\bm\sigma_j)$. Additionally, regularization, typically an $L_2$ penalty, is used.  
By optimizing these sums of log-conditional probabilities in eq.~\eqref{eq:pesudo_condi_log_likelihood}  over all sites and sequences, PLM efficiently estimates the local fields and coupling parameters, allowing for accurate inference in high-dimensional genomic datasets where exact likelihood maximization by eq.~\eqref{eq:ML-r} is computationally infeasible. 

\begin{figure}[!ht]
\centering  
\includegraphics[width=0.45\textwidth]{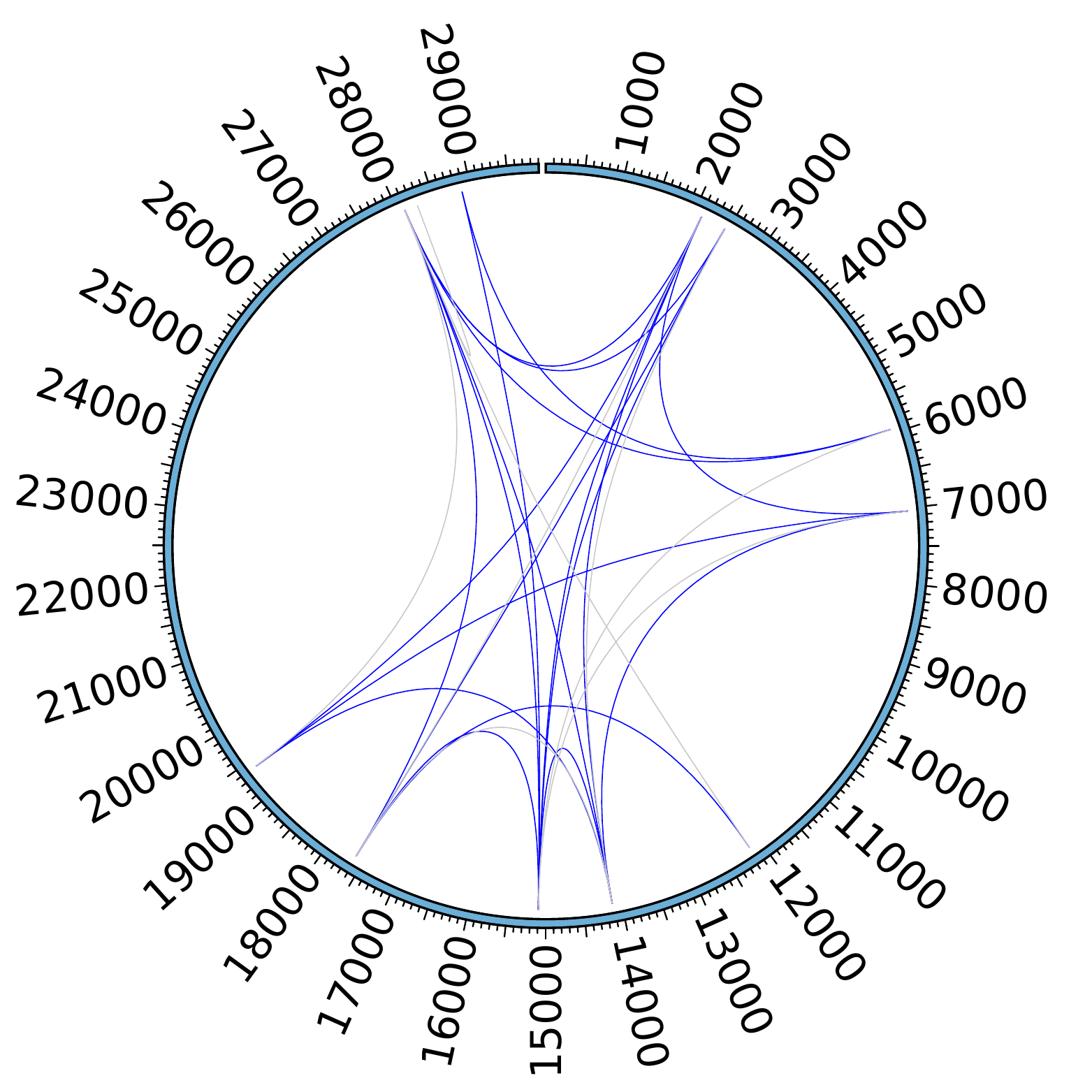}
\put(-170,175){(a)}
\includegraphics[width=0.45\textwidth]{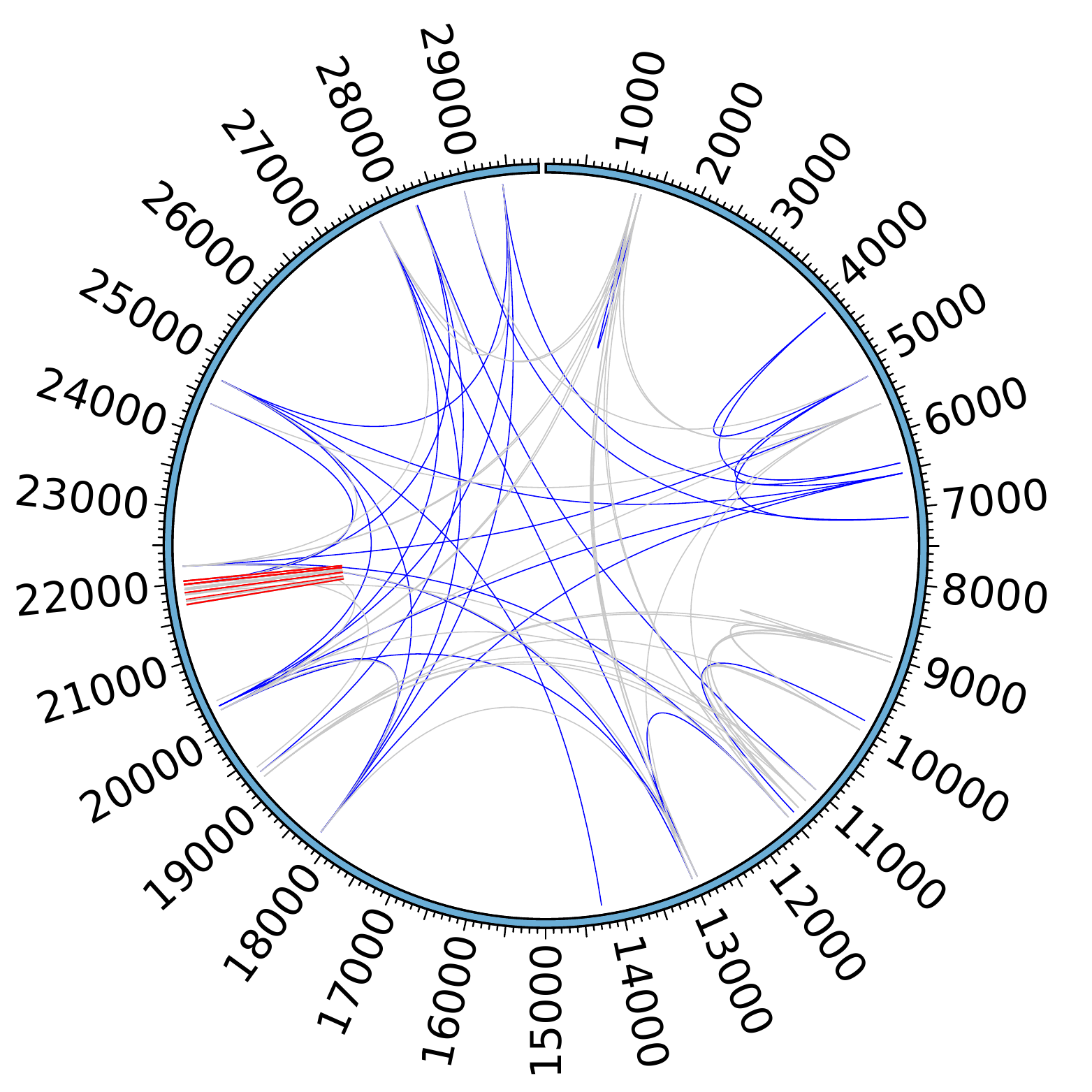}
\put(-170,175){(b)}
\\
\includegraphics[width=0.45\textwidth]{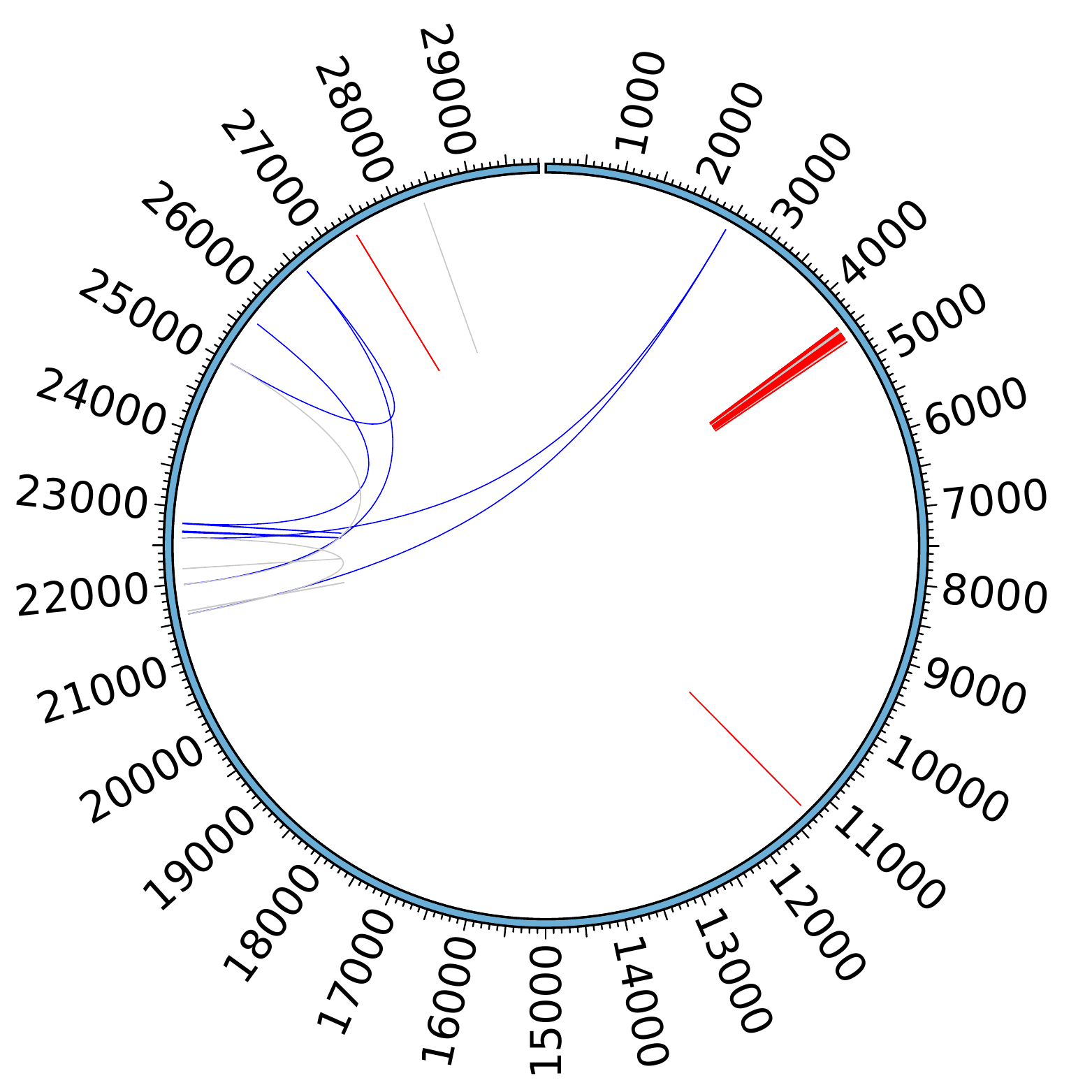}
\put(-170,175){(c)}
\includegraphics[width=0.45\textwidth]{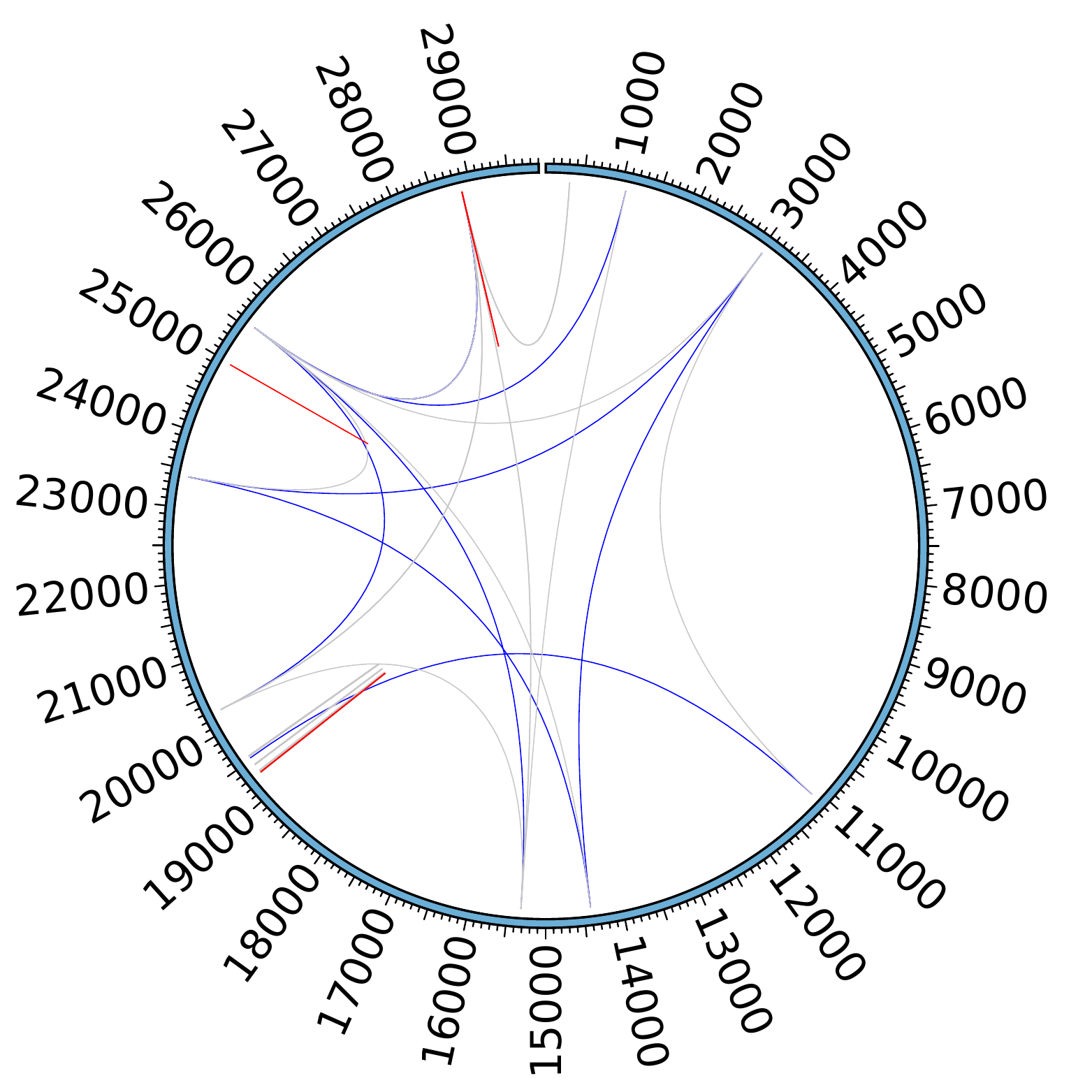}
\put(-170,175){(d)}
\caption{The graph is an image of the epistasis of data sets, with the line segments indicating the top 200 significant pairwise epistasis between the coding region motifs in the data set for this date. The colored lines indicate the top 50, and the gray lines indicate the top 51 to 200. Red lines indicate short-distance connections (distance less than or equal to 3 bp); blue lines indicate longer distances. Threshold $\phi$=95\%,(a) Alpha; (b) Delta; (c) Omicron; (d) Else.} 
\label{fig:curcos_plot} 
\end{figure}

We here take the inferred parameters $J_{ij}$ as proxies for epistatic fitness parameters. In the setting of QLE this corresponds to the low mutation rate inference formula 
$f_{ij}^*=rc_{ij}J_{ij}^*$~\cite{Kimura1965,NeherShraiman2009,NeherShraiman2011} and ignoring the overall numerical factor $r$ (recombination rate) and loci-specific factor $c_{ij}$ (related to genomic distance between $i$ and $j$).
With PLM the interaction between loci $i$ and $j$ was scored by Frobenius parametric number. 
To extract the most informative epistatic interactions from the SARS-CoV-2 genome, we selected the top 200 locus pairs with the highest coupling scores inferred by DCA. The genomic positions of the corresponding nucleotides were mapped by aligning them to the reference sequence ``Wuhan-Hu-1". These pairwise interactions were subsequently visualized using the Circos software \cite{krzywinski2009circos}, resulting in a comprehensive graphical representation of the epistatic network across the SARS-CoV-2 genome, as illustrated in Fig. \ref{fig:curcos_plot}. Plots corresponding to Beta and Gamma variants are grouped into the ``Else" category to mitigate class imbalance and ensure sufficient sample size for reliable model training.

Based on the filtering thresholds of $\phi=0.95$ and $\phi=0.96$, the number of generated epistatic interaction (Circos) images varied across different variants. To ensure balanced representation, we randomly selected approximately 500 images from each variant category, resulting in a total of 1,984  images used for training the convolutional neural networks (CNN). The whole dataset was randomly divided into 80\% for training and 10\% for validation to optimize model performance. The other 10\% is used for testing. While this ensures balanced sampling, it may introduce correlations between training and validation sets due to temporal proximity as the images are generated from temporally stratified subsets of sequences.  A detailed distribution of the images across different variants is provided in Table~\ref{tab.1}. The dataset is representative of the overall image population generated by our analysis pipeline, and similar Circos plots used for CNN training are publicly available in the GitHub repository \cite{JingBo2023}.

\newcommand{\tblwidth}{0.6\textwidth}
\begin{table}[!ht]
\caption{Summary of the dataset used for classification, showing the number of samples per class under the 80\%/10\%/10\% training – validation - testing split.} \label{tab.1}
\begin{tabular*}{\tblwidth}{@{}ccccc@{}}
\toprule
 Category & Training & Validation  & Testing & Total \\ 
\midrule
 Alpha & 415 & 52 & 51 & 518 \\
 Delta & 415 & 52 & 51 & 518 \\
 Omicron & 358 & 45 & 45 & 448 \\
 Else & 400 & 50 & 50 & 500\\
\bottomrule
\end{tabular*}
\end{table}

\subsection{Methods for image classification}

To classify the Circos plots representing SARS-CoV-2 epistatic interactions, we employed four representative convolutional neural network (CNN) architectures~\cite{li2021survey}: VGG13 and 16, MobileNetV3, DenseNet121, and EfficientNetB0. These networks embody distinct design philosophies across different generations of CNN research, allowing for a comprehensive comparison of depth, connectivity, and computational efficiency.

\subsubsection{VGG}\label{sec3.3.1}

The VGGNet family, introduced by Simonyan and Zisserman in 2014~\cite{simonyan2014very}, represents one of the most classical and widely adopted convolutional neural network (CNN) architectures for visual feature extraction. It is built on the principle of stacking small  $3\times3$ convolutional kernels with stride 1 and padding 1, followed by $2\times2$ max-pooling layers with stride 2. This design enables a deep yet computationally efficient network that progressively captures spatial hierarchies from low to high level features.

Among the VGG variants, VGG13 and VGG16 mainly differ in network depth: VGG13 contains 13 weight layers (11 convolutional and 2 fully connected), whereas VGG16 extends to 16 weight layers (13 convolutional and 3 fully connected). Both models are composed of five convolutional blocks, where the number of filters doubles after each block (64, 128, 256, 512, and 512). The extracted feature maps are flattened and passed through two fully connected layers with 4096 neurons each, followed by a Softmax layer for multi-class classification.  Generally, VGG provides a baseline for feature extraction and image classification, serving as a reference point for deeper and more efficient designs.

\subsubsection{MobileNet}\label{sec3.3.2}

MobileNets~\cite{howard2017mobilenets}, is a lightweight CNN architecture specifically designed for efficient computation on limited-resource environments while maintaining high accuracy. Unlike traditional CNNs that rely on standard convolution operations, MobileNet adopts depthwise separable convolutions to decouple spatial and channel-wise filtering. This operation drastically reduces the number of trainable parameters and computational cost, achieving a balance between accuracy and efficiency.

Here the third generation of the MobileNet family, MobileNetV3~\cite{howard2019searching}, was fine-tuned on the SARS-CoV-2 Circos image dataset using ImageNet pre-trained weights.  MobileNetV3 integrates neural architecture search (NAS), Squeeze-and-Excitation (SE) attention modules, and the hard-swish activation function to further enhance feature expressiveness and computational efficiency. These refinements enable the model to achieve high accuracy with minimal parameters, making it particularly effective for large-scale image classification under constrained resources.

\subsubsection{DenseNet}\label{sec3.3.3}

DenseNet, proposed by Huang et al. in 2017~\cite{huang2017densely}, is a deep CNN that introduces the concept of dense connectivity to enhance feature reuse and gradient propagation across layers. Unlike traditional feed-forward architectures, where each layer receives input only from its immediate predecessor, DenseNet establishes direct connections from each layer to all subsequent layers. Formally, the feature map of the $l$-th layer is defined as
$$x_l = H_l([x_0, x_1,...,x_{l-1}]),$$ where $[x_0, x_1,...,x_{l-1}]$ denotes the concatenation of feature maps produced by all previous layers, and $H_l(\cdot)$ represents a composite function consisting of Batch Normalization, ReLU activation, and $3\times3$ convolution.

The dense connectivity pattern alleviates the vanishing-gradient problem, promotes efficient feature reuse, and significantly reduces the number of parameters compared with networks of similar depth. DenseNet121 is composed of four dense blocks connected by transition layers, which perform $1\times1$ convolutions followed by $2\times2$ average pooling to control feature-map dimensions and computational cost.  DenseNet121 facilitates efficient feature reuse and robust gradient propagation, leading to improved representational capacity with fewer parameters.

\subsubsection{EfficientNet}

EfficientNetB0, proposed by Tan and Le in 2019~\cite{tan2019efficientnet}, represents a new generation of CNN optimized through compound scaling. Unlike traditional approaches that scale depth, width, or input resolution independently, EfficientNet employs a unified scaling strategy that jointly adjusts all three dimensions using a small set of fixed coefficients. This design allows the network to achieve an optimal trade-off between accuracy and computational efficiency, outperforming previous architectures with significantly fewer parameters and floating-point operations.

EfficientNetB0 is the baseline model of the EfficientNet family, discovered via neural architecture search (NAS) and subsequently scaled to larger versions (B1–B7). Its backbone structure consists of a series of mobile inverted bottleneck convolution (MBConv) blocks, which incorporate depthwise separable convolutions, squeeze-and-excitation (SE) attention modules, and the swish activation function. The MBConv design effectively reduces computational redundancy while maintaining high representational capacity, and the SE blocks dynamically recalibrate feature responses to emphasize informative channels.

The above models are fine-tuned under identical training settings as detailed in the following Section, ensuring consistent optimization and fair comparison across all employed models.

\begin{table*}[!ht]
\centering 
\caption{Hyperparameter settings for different network models}\label{tab:2}
\begin{tabular}{P{2.15cm}P{0.8cm}P{1.4cm}P{0.8cm}P{2.0cm}P{1.5cm}P{1.3cm}}
\toprule
 Model&Epoch& Activation & Batch& Loss&Optimizer& Learning  \\ 
 &  & function & Size & function &  & rate \\ 
\midrule
 VGG-13 & 100 & Softmax & 16 & CrossEntropy & SGD & 0.00625 \\
 VGG-16 & 100 & Softmax & 16 & CrossEntropy & SGD & 0.00625 \\
 MobileNetV3 & 100 & Softmax & 16 & CrossEntropy & SGD & 0.00625 \\
 DenseNet121 & 100 & Softmax & 16 & CrossEntropy & SGD & 0.00625\\
 EfficientNet\_B0 & 100 & Softmax & 16 & CrossEntropy & SGD & 0.00625\\
\bottomrule
\end{tabular}
\end{table*}

\subsubsection{Model Training Configuration}  \label{sec:training}

The setting of hyperparameters plays a crucial role in determining the performance of deep learning models. Therefore, a series of experiments were conducted to select the most suitable hyperparameter configuration for all CNN architectures. Specifically, the learning rate, batch size, and number of iterations were systematically adjusted to identify the optimal setup. The final hyperparameter settings are summarized in Table~\ref{tab:2}.

All models were implemented using the TensorFlow 2.12 framework with GPU acceleration and were initialized with ImageNet pre-trained weights. The input Circos images were uniformly resized to $224 \times 224 \times 3$, normalized to the $[0,1]$ range, and augmented with random rotation, horizontal flipping, and brightness adjustment to improve generalization. The stochastic gradient descent (SGD) optimizer with a momentum of 0.9 was employed, and the initial learning rate was set to $0.003125$ while $0.00625$ for the EfficientNet\_B0. The learning rate was decayed by a factor of 0.1 every 30 epochs (i.e., at epochs 30, 60, and 90).

The models were trained using a batch size of 16 for up to 100 epochs, and the categorical cross entropy loss function was used for multi-class classification. To prevent overfitting and optimize computational resource utilization, early stopping was applied when the  training loss failed to improve for 10 consecutive epochs. In addition, the training process was automatically suspended when the loss value stabilized, preventing unnecessary cost over computational resources. Each experiment was three times with different random seeds, and the reported results represent the mean performance across runs to ensure reproducibility and robustness.

\begin{figure}[!ht]
\centering 
\includegraphics[width=0.75\textwidth]{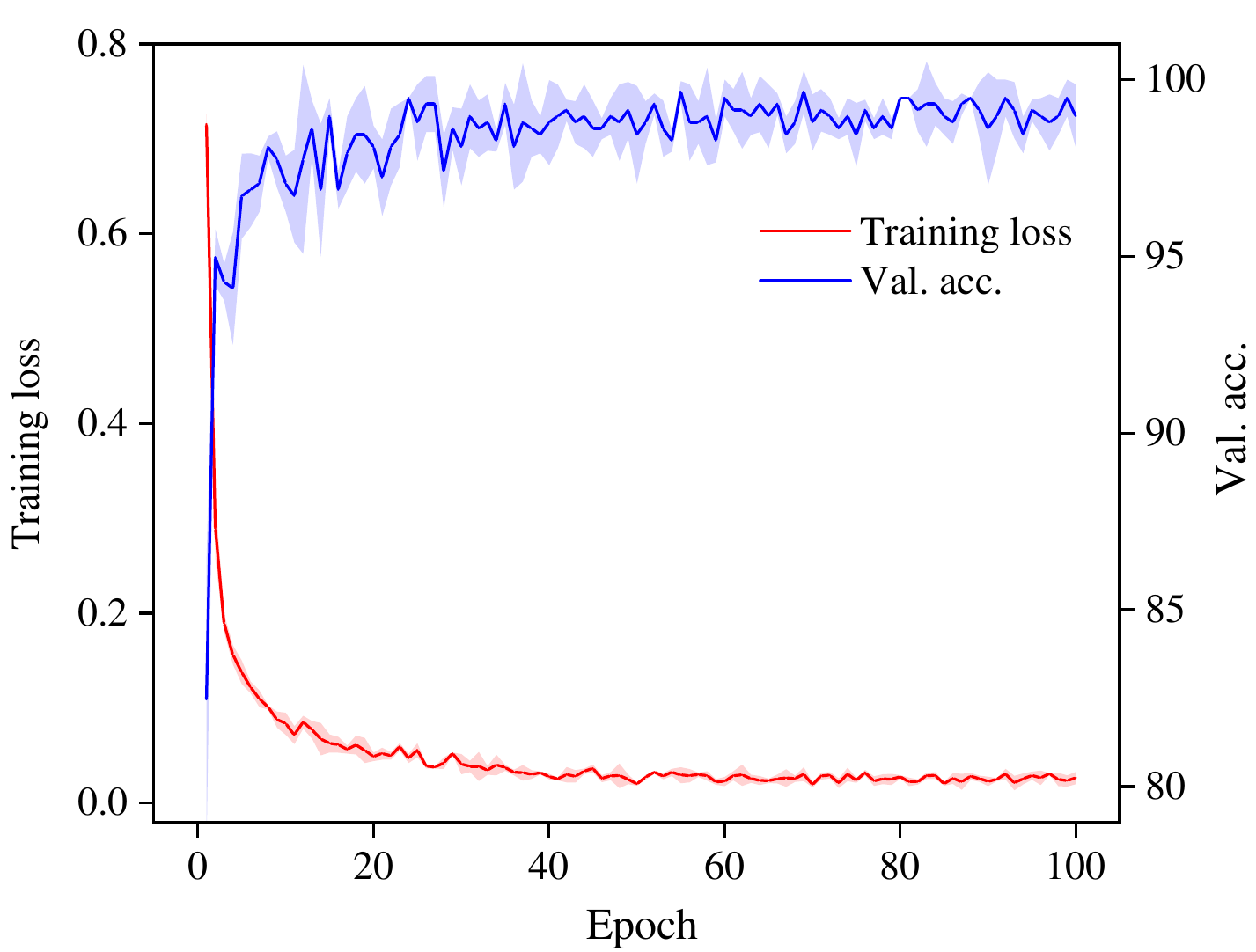} 
\caption{Training loss and validation accuracy curves of the DenseNet121 model over 100 epochs.  The training loss decreases steadily while the validation accuracy remains consistently high, indicating stable convergence and strong generalization without overfitting. Here the curves are mean values averaged over three independent runs.}
\label{fig:loss_and_accuration}
\end{figure}

To illustrate the training and convergence behavior of the models, the training dynamics of DenseNet121 as a representative example were shown in fig.~\ref{fig:loss_and_accuration}. The training loss (red curve) decreases rapidly during the initial 20 epochs and gradually stabilizes near zero, while the validation accuracy (blue curve) remains consistently above $98\%$, indicating efficient optimization and robust generalization without overfitting. 
Such behaviors benefit from the dense connectivity of DenseNet, which enhances gradient flow and feature reuse across layers, resulting in faster convergence and reduced parameter redundancy. The stable learning pattern further confirms that the chosen training strategy and hyperparameter configuration in Table~\ref{tab:2} were appropriate for the classification of SARS-CoV-2 epistatic interaction images. These convergence characteristics are consistent across other CNN architectures and provide a solid foundation for the subsequent comparative performance analysis.
Here both training and  validation performance are monitored. The training process was not terminated until the fixed number of epochs arrived. In our cases, both validation accuracy and training loss become stable with 100 epochs.  The final model was selected as the checkpoint achieving the best validation accuracy.

\section{Model Evaluation and Performance Analysis}\label{sec:results}

To comprehensively assess model performance and generalization capability, five CNN architectures, VGG16, MobileNetV3, MobileNetV3, DenseNet121, and EfficientNet\_B0, were trained and validated on the SARS-CoV-2 Circos image dataset. All models were optimized under identical hyperparameter configurations described in Table~\ref{tab:2} to ensure a fair and consistent comparison across architectures. 

Model performance was quantitatively evaluated using standard classification metrics, including accuracy, precision, recall, $F_1$-score, and the area under the receiver operating characteristic curve (AUC-ROC), as defined in the following formulae. 

\begin{align}
\text{Accuracy} &=\dfrac{TP+TN}{TP+FP+TN+FN},\\
\text{Recall} &=\dfrac{TP}{TP+FN},\\
\text{Precision} &=\dfrac{TP}{TP+FP},\\
F_1\text{-score} &= 2 \times \dfrac{\text{Precision} \times \text{Recall}}{\text{Precision} + \text{Recall}}
\end{align}
where $TP$, $TN$, $FP$, and $FN$ denote the numbers of true positives, true negatives, false positives, and false negatives, respectively. 
\text{Accuracy} measures the overall proportion of correctly classified samples and reflects the model’s general predictive capability. 
\text{Recall}, or sensitivity, quantifies the proportion of actual positive samples correctly identified by the model, indicating its ability to minimize false negatives. 
\text{Precision} evaluates the proportion of predicted positive samples that are truly positive, representing the reliability of positive predictions and the ability to reduce false alarms. 
The \text{$F_1$-score}, defined as the harmonic mean of precision and recall, provides a balanced measure of both sensitivity and specificity, particularly useful when class distributions are imbalanced. 
Additionally, the \text{AUC-ROC} (area under the receiver operating characteristic curve) was computed to assess discriminative capability of a model across varying decision thresholds. 
A higher AUC indicates a stronger separation between positive and negative classes.

\subsection{Performance from scratch}
Table \ref{tab:3} summarizes the experimental results of the five CNN architectures trained on SARS-CoV-2 Circos image datasets corresponding to different viral variants. Results are reported as mean $\pm$ standard deviation over three independent runs. Overall, all models achieved high classification accuracy, demonstrating that convolutional architectures can effectively capture the epistatic patterns embedded in the genomic interaction images.

Among the classical VGG networks, VGG-13 slightly outperformed VGG-16, achieving a weighted average $F_1$-score of $89.68 \pm2.41\%$ compared with $88.88 \pm0.73\%$ for VGG-16. The deeper structure of VGG-16 likely introduced additional parameters without providing proportional performance gains, leading to mild overfitting.

In contrast, lightweight models such as MobileNetV3 exhibited significantly improved generalization, with a weighted $F_1$-score of $ 95.98 \pm 1.83\%$  and an AUC of $0.99811\pm 0.00126$, reflecting both high accuracy and computational efficiency. Similarly, DenseNet121 achieved strong and stable results, yielding a weighted $F_1$-score of  $96.26 \pm 0.87\%$ and an AUC of $0.9989\pm0.00019$, confirming the benefit of dense connectivity in feature propagation and gradient reuse.

The best overall performance was achieved by EfficientNet\_B0, which reached a weighted $F_1$-score of  $97.3 \pm 1.22\%$ and an AUC of $0.9994 \pm 0.00014$. Its compound scaling strategy effectively balances network depth, width, and resolution, enabling superior feature extraction with fewer parameters. The consistency between $F_1$-score and AUC across all architectures indicates robust classification of variant-specific genomic patterns.

\begin{table*}[!ht]
\caption{Performance comparison of five CNN architectures on SARS-CoV-2 Circos image datasets corresponding to different viral variants.}\label{tab:3}
\begin{tabular}{P{2.2cm}P{2.0cm}P{1.8cm}P{1.8cm}P{1.8cm}P{1.2cm}}
\toprule
Models & Class & Precision & Recall & \text{$F_1$}-score & AUC \\ 
\midrule
\multirow{5}{*}{VGG-13} & Alpha & $91.74\pm3.53$ & $92.16\pm3.92$ & $91.85\pm0.60$ & \multirow{5}{*}{\shortstack{$0.98776$\\$\pm 0.0035$}}\\
 & Delta & $92.86\pm2.73$ & $92.16\pm7.07$ & $92.38\pm3.34$\\
 & Omicron & $84.85\pm7.66$ & $91.67\pm6.94$ & $88.04\pm6.60$\\
 & Else & $90.05\pm2.59$ & $83.33\pm5.77$ & $86.45\pm2.48$\\ 
 & \cellcolor{gray!20}WeightedAvg. &  \cellcolor{gray!20}$89.87\pm2.41$ &  \cellcolor{gray!20}$89.83\pm2.47$ & \cellcolor{gray!20}$89.68\pm2.41$ & \\
 \midrule
 \multirow{5}{*}{VGG-16} & Alpha & $89.76\pm3.58$ & $84.97\pm4.08$ & $87.24\pm2.63$ & \multirow{5}{*} {\shortstack{$0.98521$\\$\pm 0.0020$}}\\
 & Delta & $91.24\pm3.52$ & $94.12\pm1.96$ & $92.62\pm1.87$\\
 & Omicron & $84.24\pm2.02$ & $92.42\pm5.25$ & $88.05\pm1.53$\\
 & Else & $90.89\pm4.78$ & $84.67\pm2.31$ & $87.62\pm2.53$\\ 
 & \cellcolor{gray!20}WeightedAvg. & \cellcolor{gray!20}$89.03\pm0.80$ & \cellcolor{gray!20}$89.04\pm0.79$ & \cellcolor{gray!20}$88.88\pm0.73$ & \\
 \midrule
 \multirow{5}{*}{MobileNetV3} & Alpha & $93.56\pm2.20$ & $94.77\pm4.08$ & $94.14\pm2.67$ & \multirow{5}{*}{\shortstack{$0.99811$\\$\pm 0.00126$}}\\
 & Delta & $98.67\pm2.31$ & $96.73\pm2.26$ & $97.69\pm2.29$\\
 & Omicron & $97.76\pm2.22$ & $97.73\pm0$ & $97.74\pm1.11$\\
 & Else & $94.09\pm3.35$ & $94.67\pm1.15$ & $94.37\pm2.25$\\ 
 & \cellcolor{gray!20}WeightedAvg. & \cellcolor{gray!20}$96.02\pm1.90$ & \cellcolor{gray!20}$95.97\pm1.77$ & \cellcolor{gray!20}$95.98\pm1.83$ & \\
 \midrule
 \multirow{5}{*}{DenseNet121} & Alpha & $96.76\pm3.98$ & $94.12\pm0$ & $95.39\pm1.96$ & \multirow{5}{*}{\shortstack{$0.9989$ \\ $\pm 0.00019$}}\\
 & Delta & $98.06\pm1.89$ & $97.39\pm2.99$ & $97.70\pm1.54$\\
 & Omicron & $96.98\pm1.16$ & $96.21\pm6.56$ & $96.50\pm3.21$\\
 & Else & $93.61\pm2.14$ & $97.33\pm1.15$ & $95.43\pm1.48$\\ 
 & \cellcolor{gray!20}WeightedAvg. & \cellcolor{gray!20}$96.35\pm0.75$ & \cellcolor{gray!20}$96.26\pm0.96$ & \cellcolor{gray!20}$96.26\pm0.87$ &  \\
 \midrule
 \multirow{5}{*}{EfficientNet\_B0} & Alpha & $96.86\pm3.83$ & $98.04\pm1.96$ & $97.42\pm2.40$ & \multirow{5}{*}{\shortstack{$0.9994$ \\ $\pm0.00014$}}\\
 & Delta & $98.10\pm1.89$ & $99.34\pm1.13$ & $98.70\pm0.54$\\
 & Omicron & $99.26\pm1.28$ & $96.21\pm4.73$ & $97.66\pm2.09$\\
 & Else & $96.76\pm3.98$ & $96.67\pm1.15$ & $96.69\pm2.45$\\ 
 & \cellcolor{gray!20}WeightedAvg. & \cellcolor{gray!20}$97.43\pm1.04$ & \cellcolor{gray!20}$97.27\pm1.31$ & \cellcolor{gray!20}$97.30\pm1.22$ &  \\
\bottomrule
\end{tabular}
\end{table*}

\subsection{Performance from transfer learning}
The classification results in Table \ref{tab:3} were obtained from models trained with randomly initialized weights. To enhance convergence speed and generalization performance, we subsequently fine-tuned the same architectures using pre-trained weights derived from the ImageNet dataset. In transfer learning, low- and mid-level visual features such as edges, textures, and shapes, which are learned from large-scale natural image corpora like ImageNet, are generally transferable across tasks. To balance the contribution of different feature levels, lower convolutional layers are typically frozen to retain general visual representations, while higher layers are fine-tuned to adapt to task-specific semantics. This hierarchical adjustment allows the model to leverage generic visual priors while maintaining flexibility for the new classification of the SARS-CoV-2 Circos image dataset here. 
Meanwhile, the learning rate was reduced by an order of magnitude relative to the initial training (e.g., from $3.125 \times 10^{-3}$ to $3.125 \times 10^{-4}$) to prevent the degradation of pre-trained features. Data augmentation techniques such as random rotation, horizontal flipping, and scaling were again applied to improve model generalization.

The transfer learning strategy led to faster convergence and higher stability across all architectures, as the models required fewer epochs to reach optimal validation accuracy.
In particular, EfficientNet\_B0 and DenseNet121 exhibited notable gains in both $F_1$-score and AUC,  demonstrating that pre-trained weights significantly facilitate the extraction of complex visual features associated with these variants.

Although Circos plots are synthetic genomic visualizations structurally distinct from natural images, these observed gains likely stem from the extraction of generic geometric features (e.g., edges and contours) in early layers, coupled with improved initialization and implicit regularization. However, it is important to note that these improvements reflect general visual optimization rather than the learning of underlying biological mechanisms. Therefore, exploring domain specific pretraining, such as self-supervised learning on genomic visualizations or sequence native representations remains a critical future direction to capture deeper, biologically meaningful structures.

\begin{table*}[!ht]
\caption{Experimental results of five models using transfer learning with different virulent strain data samples}\label{tab:4}
\begin{tabular}{P{2.2cm}P{2.0cm}P{1.8cm}P{1.8cm}P{1.8cm}P{1.2cm}}
\toprule
 Models & Class & Precision & Recall & F1-score & AUC \\ 
\midrule
\multirow{5}{*}{VGG-13} & Alpha & $98.03\pm0.02$ & $97.39\pm1.13$ & $97.70\pm0.58$ & ~~
\multirow{5}{*}{\shortstack{$0.99934$ \\ $\pm 0.00081$}} \\
 & Delta & $100\pm0$ & $98.69\pm1.13$ & $99.34\pm0.57$\\
 & Omicron & $97.13\pm3.24$ & $100\pm0$ & $98.53\pm1.68$\\
 & Else & $97.99\pm0.05$ & $97.33\pm2.31$ & $97.65\pm1.18$\\ 
 & \cellcolor{gray!20}WeightedAvg. & \cellcolor{gray!20}$98.28\pm0.83$ & \cellcolor{gray!20}$98.35\pm0.76$ & \cellcolor{gray!20}$98.30\pm0.80$ &\\
 \midrule
 \multirow{5}{*}{VGG-16} & Alpha & $96.1\pm0.16$ & $96.73\pm4.08$ & $96.39\pm2.13$ & \multirow{5}{*}{\shortstack{$0.99896$ \\ $\pm 0.00018$}} \\
 & Delta & $98.11\pm3.27$ & $98.04\pm0$ & $98.06\pm1.65$\\
 & Omicron & $97.75\pm0.03$ & $98.49\pm1.31$ & $98.11\pm0.66$\\
 & Else & $99.33\pm1.15$ & $98\pm0$ & $98.66\pm0.57$\\
 & \cellcolor{gray!20}WeightedAvg. & \cellcolor{gray!20}$97.82\pm0.75$ & \cellcolor{gray!20}$97.81\pm0.70$ & \cellcolor{gray!20}$97.81\pm0.73$ & \\
 \midrule
 \multirow{5}{*}{MobileNetV3} & Alpha & $99.33\pm1.15$ & $96.73\pm1.13$ & $98.01\pm0.99$ & \multirow{5}{*}{\shortstack{$0.99974$ \\ $\pm 0.00022$}}\\
 & Delta & $100\pm0$ & $98.69\pm1.13$ & $99.34\pm0.57$\\
 & Omicron & $99.26\pm1.28$ & $99.24\pm1.31$ & $99.24\pm0.66$\\
 & Else & $96.18\pm1.85$ & $100\pm0$ & $98.05\pm0.96$\\
 & \cellcolor{gray!20}WeightedAvg. & \cellcolor{gray!20}$98.69\pm0.31$ & \cellcolor{gray!20}$98.67\pm0.31$ & \cellcolor{gray!20}$98.66\pm0.30$ &  \\
 \midrule
 \multirow{5}{*}{DenseNet121} & Alpha & $98.72\pm2.22$ & $96.73\pm2.26$ & $97.69\pm1.14$ & \multirow{5}{*}{\shortstack{$0.99973$ \\ $\pm 0.00012$}}\\
 & Delta & $98.74\pm2.18$ & $100\pm0$ & $99.36\pm1.11$\\
 & Omicron & $100\pm0$ & $100\pm0$ & $100\pm0$\\
 & Else & $97.36\pm1.11$ & $98\pm2$ & $97.67\pm1.16$\\
 & \cellcolor{gray!20}WeightedAvg. & \cellcolor{gray!20}$98.71\pm0.73$ & \cellcolor{gray!20}$98.68\pm0.76$ & \cellcolor{gray!20}$98.68\pm0.75$ &\\
 \midrule
 \multirow{5}{*}{EfficientNet\_B0} & Alpha & $98\pm2.04$ & $96.73\pm4.08$ & $97.35\pm2.90$ & \multirow{5}{*}{\shortstack{$0.99935$ \\ $\pm 0.00052$}}\\
 & Delta & $99.36\pm1.10$ & $99.35\pm1.12$ & $99.35\pm0.57$\\
 & Omicron & $98.50\pm1.30$ & $98.49\pm1.31$ & $98.49\pm0.66$\\
 & Else & $94.75\pm2.24$ & $96\pm2$ & $95.37\pm2.05$\\
 & \cellcolor{gray!20}WeightedAvg. & \cellcolor{gray!20}$97.65\pm1.07$ & \cellcolor{gray!20}$97.64\pm0.98$ & \cellcolor{gray!20}$97.64\pm1.03$ &  \\
\bottomrule
\end{tabular}
\end{table*}

\subsubsection{Comparative Analysis between Random Initialization and Transfer Learning}

The results in  Table~\ref{tab:4}  shows the substantial performance improvement achieved through transfer learning. When initialized with pre-trained ImageNet weights, all five CNN architectures exhibited faster convergence, higher classification accuracy, and stronger generalization across variant categories. Specifically, the weighted average $F_1{text{-scores}}$ of all models improved by approximately $2$ to $3\%$, while the AUC values increased consistently, all exceeding $0.99$.
The gain was most pronounced for DenseNet121, whose weighted $F_1{text{-scores}}$ reached  $98.68 \pm 0.76\%$, marking an improvement of nearly $2.5\%$ over its randomly initialized counterparts.
These two architectures benefited the most from fine-tuning, as the dense connectivity in DenseNet facilitate efficient adaptation of pre-trained representations to the SARS-CoV-2 Circos images.

Furthermore, the improvement in VGG-13 and VGG-16 indicates that even classical CNNs gain notable advantages from transfer learning, particularly in recall and stability across smaller variant classes.
The uniformly high precision and recall values (each above $97\%$) across all architectures suggest that pre-trained weights not only accelerate optimization but also effectively prevent overfitting to specific strain-level data distributions.

\begin{table*}[!ht]
\caption{Experimental results for the five models after using the Dropout layer under the use of transfer learning}\label{tab:5}
\begin{tabular}{P{2.2cm}P{2.0cm}P{1.8cm}P{1.8cm}P{1.8cm}P{1.2cm}}
\toprule
 Models & Class & Precision & Recall & F1-score & AUC \\ 
\midrule
\multirow{5}{*}{VGG-13} & Alpha & $96.17\pm1.91$ & $98.04\pm1.96$ & $97.09\pm1.68$ & \multirow{5}{*}{\shortstack{$0.99907$ \\ $\pm0.00056$}}\\
 & Delta & $98.03\pm2.00$ & $97.39\pm2.99$ & $97.70\pm2.29$\\
 & Omicron & $97.78\pm2.22$ & $99.24\pm1.31$ & $98.50\pm1.72$\\
 & Else & $97.96\pm3.53$ & $95.33\pm3.06$ & $96.63\pm3.24$\\
 & \cellcolor{gray!20} WeightedAvg. & \cellcolor{gray!20}$97.48\pm2.24$ & \cellcolor{gray!20}$97.5\pm2.97$ & \cellcolor{gray!20} $97.48\pm2.21$\\
 \midrule
 \multirow{5}{*}{VGG-16} & Alpha & $98.71\pm1.12$ & $97.39\pm2.99$ & $98.01\pm1.03$ & \multirow{5}{*}{\shortstack{$0.99978$ \\ $\pm0.00012$}}\\
 & Delta & $99.36\pm1.11$ & $100\pm0$ & $99.68\pm0.56$\\
 & Omicron & $100\pm0$ & $99.24\pm1.31$ & $99.62\pm0.66$\\
 & Else & $97.41\pm2.24$ & $98.67\pm1.15$ & $98.02\pm0.98$\\
 & \cellcolor{gray!20} WeightedAvg. & \cellcolor{gray!20}$98.87\pm0.56$ & \cellcolor{gray!20}$98.82\pm0.59$ & \cellcolor{gray!20} $98.83\pm0.58$\\
 \midrule
 \multirow{5}{*}{MobileNetV3} & Alpha & $97.36\pm1.10$ & $96.08\pm1.96$ & $96.71\pm1.16$ & \multirow{5}{*}{\shortstack{$0.99902$ \\ $\pm0.00062$}}\\
 & Delta & $97.43\pm2.22$ & $98.69\pm1.13$ & $98.06\pm1.68$\\
 & Omicron & $100\pm0$ & $97.73\pm2.28$ & $98.84\pm1.17$\\
 & Else & $94.82\pm2.86$ & $96.67\pm2.31$ & $95.72\pm2.06$\\
 & \cellcolor{gray!20} WeightedAvg. & \cellcolor{gray!20}$97.40\pm1.39$ & \cellcolor{gray!20}$97.29\pm1.49$ & \cellcolor{gray!20}$97.33\pm1.45$\\
 \midrule
 \multirow{5}{*}{DenseNet121} & Alpha & $97.41\pm2.24$ & $96.73\pm1.13$ & $97.06\pm0.96$ & \multirow{5}{*}{\shortstack{$0.99916$ \\ $\pm0.00011$}}\\
 & Delta & $99.36\pm1.11$ & $99.35\pm1.13$ & $99.35\pm0.57$\\
 & Omicron & $98.49\pm1.31$ & $98.49\pm1.31$ & $98.49\pm1.31$\\
 & Else & $97.15\pm1.11$ & $98\pm2$ & $97.67\pm1.16$\\
 & \cellcolor{gray!20} WeightedAvg. & \cellcolor{gray!20} $98.15\pm0.80$ & \cellcolor{gray!20}$98.14\pm0.80$ & \cellcolor{gray!20}$98.14\pm0.79$\\
 \midrule
 \multirow{5}{*}{EfficientNet\_B0} & Alpha & $99.35\pm1.13$ & $96.73\pm4.08$ & $97.99\pm2.04$ & \multirow{5}{*}{\shortstack{$0.99967$ \\ $\pm0.00014$}}\\
 & Delta & $98.10\pm1.89$ & $98.69\pm2.26$ & $98.37\pm0.57$\\
 & Omicron & $99.26\pm1.28$ & $98.49\pm1.31$ & $98.86\pm0.02$\\
 & Else & $96.83\pm2.89$ & $99.33\pm1.15$ & $98.04\pm0.95$\\
 & \cellcolor{gray!20} WeightedAvg. & \cellcolor{gray!20}$98.38\pm0.59$ & \cellcolor{gray!20} $98.31\pm0.54$ & \cellcolor{gray!20} $98.31\pm0.58$\\
\bottomrule
\end{tabular}
\end{table*}

\subsection{Dropout Ablation Study}

To investigate whether additional regularization could further enhance robustness under transfer learning, we applied a Dropout layer to the classifier and re-evaluated all five CNN architectures.
As shown in Table \ref{tab:5}, however, introducing Dropout consistently decreased overall performance across all models, with weighted average $F_1{text{-scores}}$ dropping by approximately $0.5-1.5\%$ compared with the non-Dropout transfer-learning results for MobileNetV3, VGG-13 and DenseNet121 while increase $1.04\%$ and $0.69\%$ for VGG-16 and EfficientNet\_B0 respectively (Table \ref{tab:4}).

The degradation is likely due to the fact that, under transfer learning, the convolutional backbone is initialized with highly stable and well-generalized feature representations learned from large-scale datasets.
In this setting, the addition of stochastic deactivation disrupts the structured feature space inherited from the pre-trained model, weakening the discriminative capacity of the high-level representations.

Moreover, the Circos images used in this study encode global epistatic interaction patterns, where the removal of randomly selected activations can obscure informative long-range dependencies, thus leading to inferior classification performance.

These findings indicate that transfer learning already provides sufficient regularization, and that additional Dropout is not only unnecessary but may even be detrimental for SARS-CoV-2 variant classification.
Therefore, all final models in this work were trained without Dropout during fine-tuning.

\subsection{Confusion Matrix and ROC curves}

\begin{figure}[!ht]
\centering 
\includegraphics[width=0.8\textwidth]{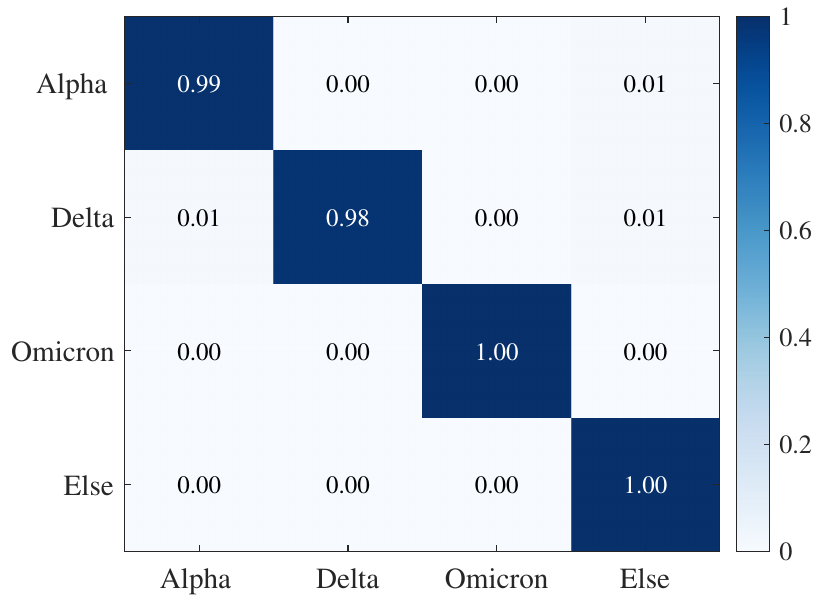}
\put(-320, 90){\rotatebox{90}{True label}}
\put(-170,-8){Predicted label}
\caption{Confusion matrix of the DenseNet121 model with transfer learning setting for predicting predicting SARS-CoV-2 variant classes from DCA-derived epistatic interaction images. Values represent normalized prediction frequencies for the four variant categories (Alpha, Delta, Omicron, and Else). The strong diagonal dominance indicates high accurate classification performance, with only minor reciprocal confusion between Alpha and Delta and perfect separation for Omicron and the composite Else class. This matrix highlights the discriminability of lineage-specific epistatic structures captured by the model.}
\label{Fig.5}
\end{figure}

The confusion matrix of the EfficientNet\_B0 classifier (fig.
~\ref{Fig.5}) exhibits a nearly ideal diagonal structure, reflecting highly consistent class-wise prediction fidelity across all four SARS-CoV-2 lineages. Misclassification events are extremely sparse and restricted to a minor degree of reciprocal confusion between Alpha and Delta, two variants known to share partially overlapping mutational profiles during specific transmission periods. In contrast, Omicron and the aggregated Else category are resolved with substantial accuracy, indicating that the model successfully captures the distinctive epistatic interaction signatures characterizing these lineages.

This pattern suggests that the convolutional features extracted from Circos-based epistasis representations contain sufficiently rich and lineage-specific structure to enable robust differentiation, even in cases where genomic backgrounds exhibit substantial evolutionary convergence. The high diagonal dominance therefore highlights not only the model’s predictive strength, but also the discriminability of epistatic architectures across major SARS-CoV-2 clades.

\begin{figure}[!ht]
\centering 
\includegraphics[width=0.7\textwidth]{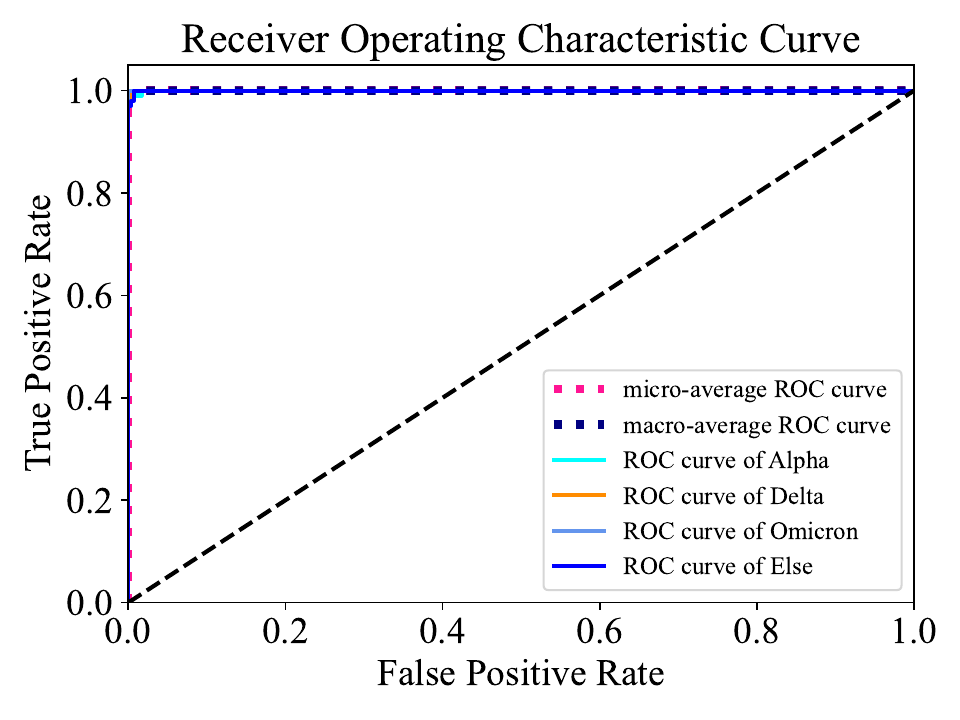}
\caption{Receiver operating characteristic (ROC) curves for the DenseNet121 classifier evaluated on the SARS-CoV-2 epistasis image dataset. Curves are shown for each variant class as well as for the micro- and macro-averaged performance across all classes. All ROC trajectories lie near the top-left corner, with area-under-the-curve (AUC) values approaching 1.0, demonstrating excellent threshold-independent discrimination and robust separability of variant-specific epistatic patterns.} 
\label{Fig.6}
\end{figure}

Moreover, the ROC curves in fig.~\ref{Fig.6} further substantiate the classifier’s exceptional discriminative capacity. All variant-specific ROC trajectories converge toward the top-left corner, and the near-perfect overlap between the micro- and macro-averaged curves indicates uniformly strong performance across both frequent and less abundant classes. AUC values approaching 1.0 demonstrate that the learned feature embeddings provide a highly stable decision boundary, insensitive to threshold variation and free from class-imbalance degradation.

These results imply that the Circos epistasis used as model inputs encode global interaction patterns that remain separable under a broad range of classification thresholds. Such robustness is particularly noteworthy in the context of rapidly evolving viral genomes, where shifts in allele frequencies and combinatorial interactions could otherwise obscure lineage boundaries. The ROC analysis therefore reinforces the conclusion that the model captures biologically meaningful epistatic structures with high generalization capability.

\begin{figure}[!ht]
\centering 
\includegraphics[width=0.8\textwidth]{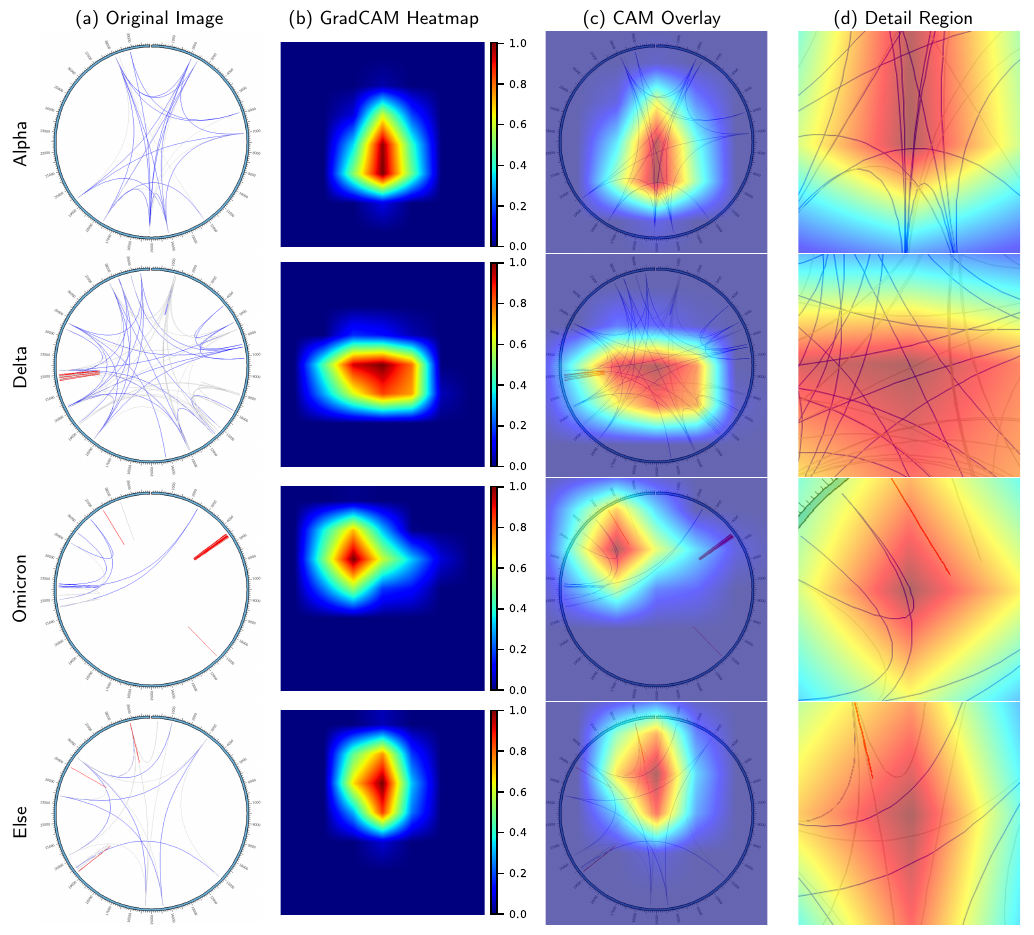}
\caption{Visualization of EfficientNet\_B0  decision-making regions for SARS-CoV-2 strain classification using GradCAM. The figure illustrates the model's attention mechanisms across four distinct classes: Alpha, Delta, Omicron, and Else. (a) The original input images representing genomic data as Circos plots. (b) The generated GradCAM heatmaps, where red regions indicate high activation values and blue regions represent low activation. (c) The overlay of the heatmap onto the original image, highlighting the specific structural features contributing to the classification. (d) A magnified view of the most salient regions. The visualization reveals that the model learns to identify strain-specific topological patterns and linkage densities within the genomic interaction networks to distinguish between variants.} 
\label{Fig.7}
\end{figure}

To validate the interpretability of the EfficientNet\_B0 model and understand the morphological features driving the classification of SARS-CoV-2 variants, we employed Gradient-weighted Class Activation Mapping (GradCAM). As shown in fig. \ref{Fig.7}, the model demonstrates distinct attention patterns corresponding to the unique genomic mutation linkages of each strain. For the Alpha variant, the ``heat" focus is prominently concentrated in the central region of the Circos plot, suggesting a reliance on a specific convergence of mutation links. In contrast, the Delta variant exhibits a broader, horizontally distributed activation region slightly offset to the left, capturing a wider network of genomic interactions. notably, the Omicron variant triggers a highly localized activation in the upper-left quadrant, distinguishing it from the central focus observed in the ``Else" category. These visualizations  indicates that the classifier does not rely on background artifacts or peripheral noise (such as the outer scale ring); instead, it highlights the discriminative topological features, specifically the intersection and density of internal chords representing mutation relationships, to achieve robust classification.

\section{Discussion}\label{sec:discussion}

The evolution of SARS-CoV-2 has been characterized by the recurrent emergence of highly divergent lineages with distinct phenotypic profiles, including altered transmissibility, immune escape potential, and host-range adaptation. Understanding the genomic determinants of these phenotypic shifts requires analytical approaches that can detect not only individual mutations but also the coordinated mutational processes that shape viral adaptation. Our study demonstrates that epistatic interaction patterns inferred from large-scale SARS-CoV-2 sequences contain rich biological information about how variants diversify and adapt, and that these patterns are sufficiently distinctive to enable robust lineage classification using deep learning.

The inferred coupling patterns and visualization results suggest structured relationships within the data, but it does not directly validate these patterns against experimentally established epistatic mechanisms. The predictive success of the models may reflect its ability to capture statistical regularities in the dataset, which could not be necessarily equivalent to causal biological interactions. Future work incorporating experimental validation or independent biological benchmarks will be necessary to establish the biological relevance of the inferred patterns.

Random splits over temporally stratified data may lead to optimistic performance estimates due to residual correlations between samples. Therefore, results obtained under stricter splits (e.g., temporal holdout) provide a more realistic assessment of generalization. While performance decreases under these settings, the model retains meaningful predictive capability, suggesting that it captures non-trivial patterns beyond local temporal correlations.

The strong lineage-specific separability observed in our epistatic maps reflects the fact that SARS-CoV-2 evolution is highly modular. Major variants such as Alpha, Delta, and Omicron exhibit characteristic constellations of mutations that arise in the context of strong selective pressures—particularly immune evasion and enhanced receptor binding. These constellations are not random collections of substitutions; rather, they represent co-evolving mutational modules shaped by both functional requirements and structural constraints of the viral proteome. For example, Omicron’s extensive set of Spike mutations includes both destabilizing and compensatory changes, forming tightly integrated epistatic networks that allow the lineage to escape neutralizing antibodies while maintaining receptor affinity. 
The DCA-inferred couplings recovered in our studies show patterns that are qualitatively consistent with previously reported biological interactions, suggesting that the statistical signatures captured by our method may reflect underlying co-adaptive mechanisms in the viral genome.

The rare misclassification between Alpha and Delta further reinforces the biological realism of our findings. Although these lineages are phylogenetically distant, both experienced phases of rapid global spread and independently acquired recurrent mutations in genomic regions under strong selective pressure, such as Spike NTD and RBD residues implicated in immune evasion. These episodes of convergent evolution create partial overlap in their epistatic profiles, explaining the slight ambiguity observed in the model’s predictions. Such patterns underscore the importance of analyzing SARS-CoV-2 evolution not solely through phylogenetic divergence but also through the lens of shared adaptive pathways and repeated mutational solutions.

The results highlight that the epistatic landscape of SARS-CoV-2 is highly structured, with major lineages occupying distinct regions of the underlying fitness landscape. This structure arises from evolutionary constraints imposed by protein stability, replication machinery fidelity, host receptor usage, and immune selection. The fact that these constraints manifest so clearly in DCA-inferred couplings suggests that pairwise epistasis captures a substantial portion of the selective architecture governing SARS-CoV-2 evolution. The ability of CNNs to learn higher-order patterns from these couplings implies that even more complex interactions—beyond pairwise correlations—may be encoded in the Circos-derived epistatic geometries.

The findings here suggest that lineage-specific epistatic signatures may serve as predictive markers of evolutionary potential, offering insight into which mutational combinations are selectively favored or disfavored under different epidemiological conditions. As new variants continue to emerge, the integrative framework presented here could support early detection of unusual epistatic configurations that signal shifts in antigenicity, host adaptation, or transmission patterns. More broadly, this work demonstrates that combining population-genetic theory, coevolutionary inference, and deep learning provides a powerful approach for exploring the complex co-adaptive landscapes of viral evolution.

\section{Conclusion}\label{sec:conclusion}

We developed an integrated framework that combines population-genetic theory, statistical epistasis inference, and deep learning to classify SARS-CoV-2 genomic variants through their epistatic interaction signatures in this study. By applying DCA approach under the QLE approximation, we derived lineage-specific coupling patterns that capture coordinated mutational dependencies shaped by the virus’s evolutionary history. Circos-based visualizations of these epistatic networks provided a compact and interpretable representation of the global interaction structure, from which CNNs were able to extract robust discriminative features.

Across multiple architectures, transfer learning markedly enhanced classification accuracy, with DenseNet121 and EfficientNet\_B0 achieving highly competitive performance. Confusion matrix and ROC analyses further confirmed the strong separability of epistatic landscapes across SARS-CoV-2 variants, indicating that the major lineages occupy distinct regions of the fitness landscape shaped by co-evolving mutational modules. These results validate not only the predictive capability of the proposed approach but also the biological relevance of DCA-inferred epistatic couplings as a meaningful descriptor of viral evolutionary constraints.

The methodological framework introduced here offers several promising directions for future genomic surveillance and evolutionary inference. First, the ability of deep networks to detect lineage-specific epistatic signatures from DCA-derived coupling maps suggests that this approach may be extended to emerging variants in real time, even before large-scale phylogenetic characterization is available. By continuously updating MSAs and recalculating coupling structures, the system could provide an early-warning signal when novel epistatic patterns deviate from established lineages, potentially indicating shifts in transmissibility, immune escape potential, or host adaptation.

Second, because DCA captures pairwise epistasis but CNNs can implicitly learn higher-order dependencies, this hybrid framework is well positioned to reveal complex interaction modules that underlie major adaptive transitions. Such modules may correspond to compensatory mutational clusters or structural integration motifs that define evolutionary accessible trajectories. Identifying these modules could enhance our understanding of the constraints shaping SARS-CoV-2 evolution and support predictive modeling of future lineage emergence.

Finally, the generality of the approach makes it directly applicable to other rapidly evolving pathogens—including influenza, HIV, and arboviruses—where coordinated evolutionary processes play a central role in antigenic drift and escape. By integrating statistical physics–based inference with modern computer vision, the framework bridges mechanistic modeling and data-driven prediction, providing a powerful tool for real-time genomic monitoring, evolutionary forecasting, and functional interpretation of mutational landscapes.

\backmatter
\section*{Data Availability}
All prepared circos plots used here can be downloaded through the link:\\
``https://github.com/xiaohuolongx/circos-data-set.git" \\
However, the sequences used to get the Circos plots are not enclosed as they can be collected from the public dataset GISAID: https://app1.epicov.org/epi3/
Codes as well as configuration files used for this manuscript are enclosed on GitHub, including the filtering script for the sequences. 
They are reachable through the following link:\\ https://github.com/xiaohuolongx/paper-Machine-Learning-SARS-CoV-2-Classification-main.git\\

\section*{Acknowledgements}

The work of BJ, HLZ was sponsored by the National Natural Science Foundation of China (11705097), the China Scholarship Council (202508320441), and the Natural Science Foundation of Nanjing University of Posts and Telecommunications (Grant No. 221101, 222134). EA acknowledges support from the Swedish Research Council through grant 2020-04980.

\section*{Declarations}

There is no Conflict of interest.

\bibliography{SARS-CoV-2-CNN}
\end{document}